\documentclass[1p]{elsarticle}

\usepackage{lineno,hyperref}
\hypersetup{
    colorlinks=true,
    linkcolor=blue,
    filecolor=magenta,      
    urlcolor=cyan,
}
\usepackage{subfig}
\usepackage{amsmath}
\usepackage{amssymb}
\usepackage{booktabs}
\usepackage{multirow}
\usepackage{multicol}
\usepackage{pgfplots}
\pgfplotsset{width=6cm,compat=1.14}
\usepackage{float}
\usepackage{rotating}
\usepackage{array,makecell}
\usepackage[export]{adjustbox}
\usepackage{arydshln}
\usepackage{graphics}
\usepackage{xspace}
\usepackage{graphicx}
\usepackage{caption}
\usepackage{soul}
\usepackage[ruled,vlined,linesnumbered]{algorithm2e}

\newcommand{\modelname}{SUCP\xspace}

\journal{Information Processing \& Management}

\usepackage{numcompress}

\begin{document}

\begin{frontmatter}

\title{Leveraging Social Influence based on Users Activity Centers for Point-of-Interest Recommendation}

\author[znu-address]{Kosar Seyedhoseinzadeh}
\ead{kosar.seyedhosseinzadeh@znu.ac.ir}

\author[ucl-address]{Hossein A. Rahmani\fnref{work}}
\ead{h.rahmani@ucl.ac.uk}

\fntext[work]{Work done at the University of Zanjan.}

\author[znu-address]{Mohsen Afsharchi}
\ead{afsharchi@znu.ac.ir}

\author[uva-address]{Mohammad Aliannejadi}
\ead{m.aliannejadi@uva.nl}

\address[znu-address]{Department of Computer Engineering, University of Zanjan, Iran}
\address[ucl-address]{Department of Computer Science, University College London, United Kingdom}
\address[uva-address]{Informatics Institute, University of Amsterdam, The Netherlands}

\begin{abstract}
Recommender Systems (RSs) aim to model and predict the user preference while interacting with items, such as Points of Interest (POIs). These systems face several challenges, such as data sparsity, limiting their effectiveness. In this paper, we address this problem by incorporating social, geographical, and temporal information into the Matrix Factorization (MF) technique. To this end, we model social influence based on two factors: \textit{similarities between users in terms of common check-ins and the friendships between them}. We introduce two levels of friendship based on explicit friendship networks and high check-in overlap between users. We base our friendship algorithm on users' geographical activity centers. The results show that our proposed model outperforms the state-of-the-art on two real-world datasets. More specifically, our ablation study shows that the social model improves the performance of our proposed POI recommendation system by $31\%$ and $14\%$ on the Gowalla and Yelp datasets in terms of Precision@10, respectively. 
\end{abstract}

\begin{keyword}
Social Influence \sep Contextual Information \sep Point-of-Interest Recommendation \sep Personalization
\end{keyword}

\end{frontmatter}

\section{Introduction and Motivation} 
\label{sec:intro}
In recent years, with the advancement of technology, we have entered the digital age, an age in which social networks become part of people's lives and create both benefits and challenges. One of the most significant issues in the digital age is information overload, which causes users difficulty finding helpful information among the available options. Overcoming this problem in the lives of people and organizations is a challenge. 
Location-Based Social Networks (LBSNs) enable users to share their experience of visiting various types of venues. 
Consequently, users are even more exposed to problems that emerge from information overload as they face many options to choose from.

To solve this problem, location recommendation systems help people make better decisions. In LBSNs, such as Foursquare\footnote{https://foursquare.com/}, Gowalla, and Yelp\footnote{https://www.yelp.com/}, users can share their experiences with Points of Interests (POIs) via check-ins while sharing other pieces of information such as rating, reviews, and images. POI recommender systems aim to predict POI's that would interest the users by mining their behaviors, aiming to help them discover new interesting POIs.

Many different approaches have been suggested for the problem of POI recommendation \cite{Berjani2011,ye2011exploiting,ye2010location,rahmani2019category}. In fact, among them, methods based on Collaborative Filtering (CF) are the most successful approaches that address different problems of POI recommendation by considering various contextual information \cite{rahmani2019lglmf,manotumruksa2020contextual,DBLP:journals/tkde/AliannejadiRC20}. The basic idea of CF is that users who visit similar items would be interested in similar items in the future \cite{schafer2007collaborative,luo2013applying}. 
In CF, a user-item matrix  is used to calculate the interactions between the users and the items. One of the traditional CF-based methods is Matrix Factorization (MF) \cite{koren2008factorization} that uses the inner product between user's and item's latent matrices to obtain user's interest in the items. CF-based methods often suffer from several shortcomings, which reduce the performance of the system. These methods require extensive data, including user rating of items or implicit feedback such as check-ins data. However, in reality, little user-item data available, resulting in a sparse user-item matrix. This problem is called data sparsity. Data sparsity poses itself as one of the main challenges of recommendation systems \cite{liu2013learning}. Researchers have proposed numerous approaches to tackle this problem. A line of research leverages different contextual information to address this problem. The main assumption in such works is that the combination of auxiliary information can extract better user behavior patterns \cite{geng2019two}. Auxiliary information in LBSNs often include 
(a) content information that is about the POI such as user comments and POI pictures; (b) geographical information that defines the physical range of user activity; (c) social information, which is about relationships between users that influence their opinions; and (d) temporal information that shows when users visited locations. 
Human beings are inherently social; therefore, they can be influenced by various factors that originate from their social relationships.
The more information we have about the factors influencing the user's choice, the better we can model their behavior pattern, hence providing more appropriate recommendations.

\begin{figure}[H]
    \centering
    \includegraphics[scale=0.6]{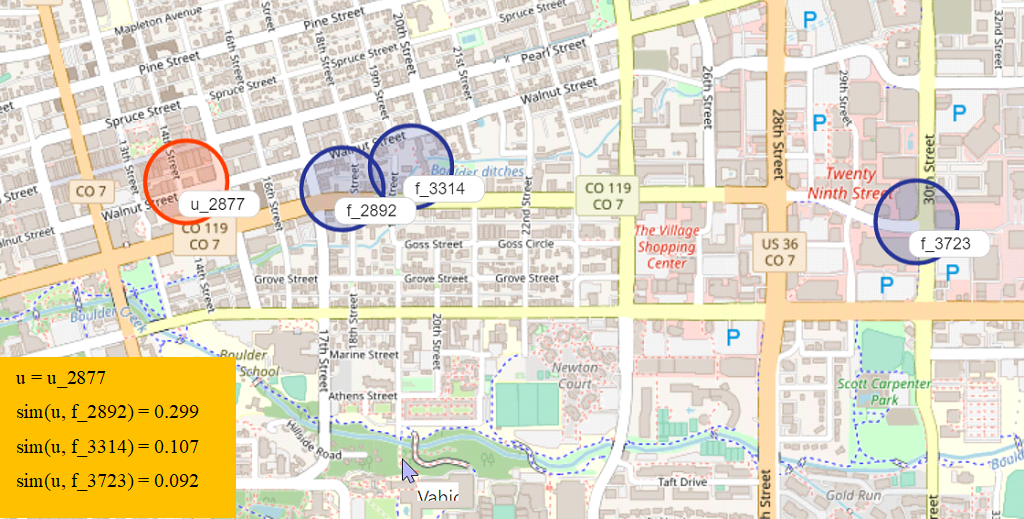}
    \caption{Friends in close neighborhood. The shorter the distance between the target user activity center (red node) and their friends (blue nodes), the greater the similarity of friendship between them.}
    \label{fig:close_dist}
\end{figure}

\begin{figure}[H]
    \centering
    \includegraphics[scale=0.6]{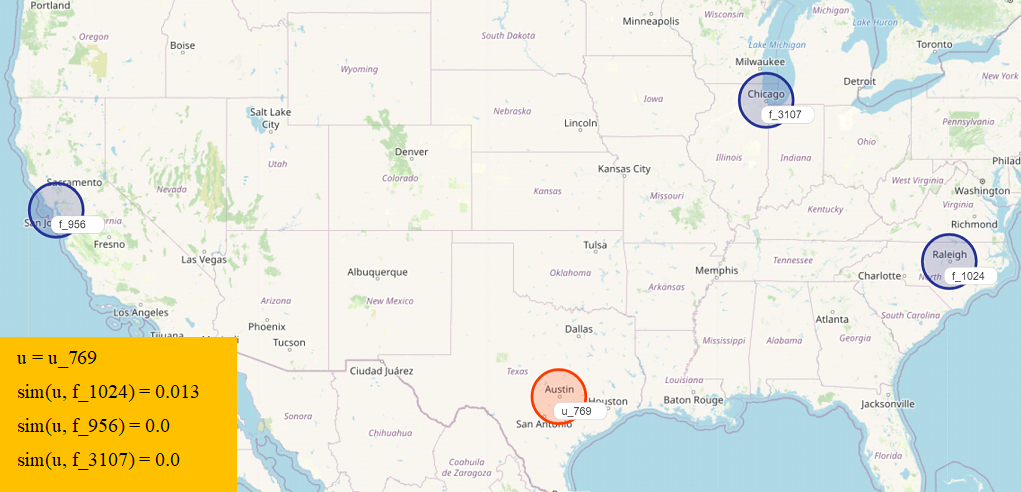}
    \caption{Friends living far from each other. The greater the distance between the target user's activity center (red node) and their friends (blue nodes), the less similarity of friendship between them.}
    \label{fig:long_dist}
\end{figure}

We define two types of friendships on LBSNs. In particular, \emph{explicit friends} are those users who follow each other in an LBSN, whereas \emph{implicit friends} are those with a high number of common check-ins who usually have very similar tastes, leading to some friendships created because of these behavioral similarities. For example, consider a user who wants to visit a location with their friends. They would ask their friends about their experiences. Such interactions also exist in social networks. Social media plays a critical role in connecting people to new people, allowing people to learn more about new people and get to know them better. Research results proposed \cite{cheng2012fused} to show that social relations limit affections on user's behaviors of check-ins, but they still should not be neglected. With this in mind, we analyze two real-world datasets based on users' social relationships to assess the impact of users' opinions on each other based on their activity centers. 
In this work, we assume that the similarity of the users is inversely proportional to their geographical distance. 
Our analysis reveals the social behavior pattern of users for geographic activity centers. We assume that the center with the most check-in number is the main user activity center, as shown in Figure \ref{fig:close_dist}. We assume that the friend of the user who has the closest activity center to the user is their most similar friend. Larger geographical distances between friends lead to lower similarity values (see Figure \ref{fig:long_dist}). 

As can be seen from the analysis, the user's geographical distance to a friend whose center of activity is close to them correlates with the level of friendship and their behavioral similarity. In this way, we consider the effects of the user activity centers assuming the similarity of their friends simultaneously and relative to each other to improve the proposed system's performance.

In the real world, we know that the distance of our desired POIs from where we are located is an essential priority of our choices. Knowing that a friend is present near that place and likes it, could affect one's interest in visiting it. In this work, we propose a \textbf{S}ocial \textbf{U}sers activity \textbf{C}enters \textbf{P}OI recommendation system, called \textbf{\modelname}, that takes the effect of social influence into account. Compared to previous work, our \modelname improves the performance of the recommendation systems by taking into account the user's social relationships in their activity centers. Our model predicts user preferences based on MF and a combination of social information and activity centers of users created under the influence of time and place.
The evaluation results demonstrate the efficiency of our proposed model. Furthermore, in order to demonstrate the effectiveness of our proposed social method, we remove the social information from \modelname ad study its effect on the performance. We see that \modelname outperforms CARA \cite{manotumruksa2018contextual} by 16\% in terms of Recall@20. Also, the results show that the proposed social model improves the performance of the POI recommendation system by $31\%$ and $14\%$ on the Gowalla and Yelp data in terms of Precision@10, respectively.
Moreover, our experiments demonstrate the effectiveness of our proposed social method. Finally, to the reproducibility of the results, we release our code as open.\footnote{\url{https://github.com/Seyedhosseinzadeh/SUCP}}

This paper is organized as follows. In Section \ref{sec:contributions}, we give our aims of this work. Then, in Section \ref{sec:relatedwork}, we review the related works and in Section \ref{sec:method} we give a detail of our proposed method. We conduct experiments in Section \ref{sec:evaluation} and the results are presented in Section \ref{sec:results}. In Section \ref{sec:discussion}, we discuss the results of our work. Finally, we conclude the paper in Section \ref{sec:conclusion}.

\section{Contributions} 
\label{sec:contributions}
This paper proposes a new approach that examines users' preferences based on three contextual factors: geographical, social, and temporal information. Although the composition of contextual information is not novel, we introduce a new approach to consider the social relationships in user's activity centers over time. We summarize the contributions of our work as follows:
\begin{itemize}
 
 \item The proposed model examines users' interest in POIs from several perspectives. The novelty of our model is to jointly model users' activity centers and their social relationships.
 \item  We consider two types of social influences including friendships and the similarities of visited POIs between users. Therefore, we achieve a weight of similarities between the user and their friends. In this way, the behavioral similarity between the user and their friends was recognized for the recommender system.
 
 \item We conduct extensive experiments on two real-world datasets to confirm the effectiveness and superiority of our \modelname. We also compare the performance of \modelname with its variant, as well as different sparsity levels. We show that \modelname outperforms competitive baselines. Also, we demonstrate the effectiveness of the social component via an ablation study.
 
\end{itemize}

\section{Related Work}
\label{sec:relatedwork}
In recent years, different studies in the domain of POI recommender systems have incorporated different contextual information into their POI recommendation models. With the popularity of LBSN, POI recommendation systems play a pivotal role in satisfying users of LBSN. CF-based methods are based on a user-POI matrix which the main idea behind that is users with similar check-in history on POIs tend to visit similar POIs in the future \cite{davtalab2020poi,ricci2002travel,ricci2015recommender,griesner2015poi}. Many research types have been done based on this idea \cite{rahmani2020mscthesis,yin2014lcars}. Although there are a lot of available locations to visit around users, users can visit and check-ins to a small number of them. This causes CF-based methods to often suffer from data sparsity. Therefore, the check-ins frequency of users and POIs becomes very sparse, and the performance of POI recommendation methods will be very poor as there are not enough check-ins in the users-POIs matrix. To this end, different studies incorporate additional information into the models to overcome this challenge \cite{adomavicius2005toward,adomavicius2011context,xia2017vrer,veras2019cd,zhu2018exploiting}.

For instance, researches on LBSNs show that users' movements have a spatial pattern. Therefore, geographical information plays an important role in LBSNs. The underlying rationale is based on \textit{Tobler's First Law of Geography}, i.e., near locations to the users are more related and available to visit than distant locations \cite{tobler1979cellular}. \citet{rahmani2019lglmf} proposed LGLMF, which is a local geographical model that incorporates geographical information from both users and location perspectives. LGLMF selects POIs that are near to the user's most visited POI from a user's perspective and applies the impact of neighbor locations from a location perspective. Then, they integrate their geographical model into an implicit MF to consider the visit frequency of users into the POIs. Also, geographical correlations among visited locations have been analyzed using a few methods to measure the range of activities \cite{zhao2013capturing,seo164point}. Although each user contains a different personalized taste for POI, the probability of visiting a POI is inversely related to the user's distance from that POI. This implies that if a place is too far away from the user's home POI, although she may like that place, she would probably not go there. \citet{cheng2012fused} have dealt with a method based on the Gaussian multi-center model in which the distinctive feature of the system is that they are located around several centers and thus show the locations close to each other as a center. They use a greedy clustering algorithm among the check-ins due to the Pareto principle \cite{hafner2001pareto} to find the centers. \citet{lian2014geomf} introduce the Weighted Matrix Factorization (WMF) method to learn the user's and POI's hidden factors using implicit feedback. Moreover, \citet{guo2019location} using WMF and geographical neighborhood present a model called L-WMF. They model geographical information as a regularization term to exploit the geographical characteristics from a POI perspective.

In addition to the geographical distance factor of places, a behavioral pattern can be extracted from the number of visits to locations by users \cite{ma2011probabilistic}. Therefore, the number of visits to the POIs is also an essential factor that shows the user's interest in that place. To this end, \citet{li2015rank} proposed a geographical factorization model which considers the user's check-in frequency and the location (i.e., latitude and longitude) of POIs to rank the POIs based on the user's preferences.

Visiting a place at the right time can make that place popular for the user. In fact, we visit different locations during the day according to various times. For example, most users prefer to go to restaurants and fast food places at noon rather than early morning. Therefore, any place at the right time can attract more visitors and be popular. Accordingly, \citet{gao2013exploring} proposed a temporal latent factorization model in POI recommendation systems. They consider the user's check-in behavior as a set of time-dependent check-in preferences, in which each preference corresponds to an hour of the day. To this end, the original user-POI matrix is divided into T sub-matrices in terms of time; then, MF is applied to these matrices one by one. Finally, the corresponding low-rank approximation is aggregated into the final matrix. Combining contextual information in location-based recommendation systems can achieve better results. This tries to examine the user's behavior from different aspects. \citet{rahmani2020joint} proposed a joint approach based on the MF that considers both geographical and temporal information jointly and is called STACP. STACP models users' activity centers based on temporal information, and they show users' centers depend on the different temporal states. \citet{manotumruksa2018contextual} proposed Contextual Attention Recurrent Architecture (CARA), a sequence-based contextual model for venue recommendation. CARA leverages both sequences of feedback and contextual information associated with the sequences to capture the users' dynamic preferences.

Inspired by the assumption which users who are friends have more common preferences than strangers users, several POI recommendation methods improve the performance of recommendation by taking social information into consideration \cite{ma2009learning,gao2018personalized,tang2016point,li2020point,xiong2020point}. \citet{zhang2013igslr} proposed a method called iGSLR that user preference, social influence, and personalized geographical influence are integrated into a unified geo-social recommendation framework. iGSLR fuses a social influence model and a geographical information-based model. To model the social relation, they consider the user's friendship similarity, and as for the geographical model, they apply a kernel density estimation approach that incorporates a two-dimensional matrix of users and location instead of a one-dimensional method. Social influence integrates the Social Collaborative Filtering (SCF) method that considers the similarity between users rather than the user's check-ins frequency as in the user-based and item-based CF methods. \citet{ye2011exploiting} proposed a geographical clustering of users' visited location activities in LBSNs based on naive bayesian to model the geographical information of locations. Moreover, they model a social-based CF approach to compute the social influence weight between two friends. They consider both the social connections and the similarity of their check-in activities. \citet{qiao2018socialmix} proposed a POI recommendation model called SocialMix that is a hybrid model that considers (i) user's familiarity and (ii) preference similarity for POI recommendation. To calculate the user's familiarity score, they use the number of mutual friends, Jaccard similarity (based on user's friend list), and cosine similarity (based on user's check-in history). The preference similarity shows users' similarity in  their preferences of visited POIs -- also calculated based on the cosine similarity of user-POI check-in data. \citet{li2016point} proposed an MF-based approach that incorporates different types of check-ins, including user's check-ins, user's friends' check-ins, and other unrelated check-ins. Then, check-ins are categorized accordingly by different types of friendships, i.e., users who follow each other, common visited location friends, and users who are geographically close to each other. Their model learns a set of potential POIs that have been visited by the user's friends, and the user may want to visit those POIs.

Moreover, \citet{zhang2015geosoca} developed GeoSoCa that includes three contextual parts, geographical, social, and categorical. GeoSoCa models geographical influence based on a kernel estimation method that computes each user check-ins distribution rather than on all user's check-ins distributions. In addition, GeoSoCa considers check-ins frequency of users' friends on a POI to compute the correlations between users and the locations. \citet{wang2013location} propose an algorithm that makes recommendations based on the user's behavior in visiting past places, the location of each POI, social relationships between users, and their similarities. They create a recommendation called LocNN based on the distance from one POI to locations previously visited by the user and calculate the similarity between users in selecting a POI using cosine similarity. Considering these two influence factors, they suggest the POI to the user using the bookmark-coloring algorithm. \citet{ma2020location} proposed a novel POI recommendation method that includes geographical, categorical, and social information with POI popularity. Geographical preferences are related to the probability of the user visiting close to their activity area. Categorical influence refers to the user's interest in certain locations by calculating the semantic similarity between POI tags. They only focused on the city level, and categorical information usually has little data and limits users to their past choices.

\section{Proposed Method}
\label{sec:method}
In this section, we describe our proposed method called \modelname. 
Our main idea is to incorporate geo-temporal activity centers of a user, as well as their friends in the recommendation process.
Such additional information increases the knowledge of the POI recommender system of the user and enables it to model their behavior from a social perspective.
The user's activity center reflects personalized interest, while the social information considers the interest from friends, which not only explicit friends but also implicit friends have an influence on user check-in behaviors.

In general, suppose $\mathcal{U} =\{u_1, u_2, u_3, ..., u_m \}$ is a collection of users and $\mathcal{L} = \{l_1, l_2, l_3, ... , l_n \}$ is a set of POIs, $m$ and $n$ are the number of users and the number of POIs, respectively. Then, the frequency of a user's check-ins in a POI can be represented by a matrix $R_{m\times{n}}$, where each cell in this matrix $r_{u,l} \in R$ represents the number of times that user $u \in \mathcal{U}$ has visited the POI $l \in \mathcal{L}$. Also, $\mathcal{L}_u$ indicates all the locations where the user $u$ visited. Finally, \modelname can be used as a general model to predict the preference of user $u$ on POI $l$ as follows:

\begin{equation}
    \modelname_{u,l} = S_{u,l} \times  TC_{u,l} \times Z_{u,l}
    \label{eq:fusion}
\end{equation}

\noindent
where $S_{u,l}$ is related to user social information, $TC_{u,l}$ the user's spatio-temporal information which considers working time and leisure time, $Z_{u,l}$ represent static and dynamic user preferences.

In user-based CF methods, a visit profile is created for each user, without considering any of the user's social relationships with other users. This profile is based on the user's check-ins to POIs and the number of check-ins, which is shown as a vector $pr_{u}$:

\begin{equation}
    pr_{u} = (w_{u,l_{1}},w_{u,l_{2}},...,w_{u,l_{|L|}})
    \label{eq:PSocial}
\end{equation}

In this vector, if $w_{u,l_{i}}$ is equal to $0$, it indicates that there is no edge between the user $u$ and the POI $l_i$. Otherwise, $w_{u,l_{i}}$ shows the number of check-ins of the user, which is also normalized as follows:

\begin{equation}
    w_ {u, l_ {i}} = freq{(u,l_ {i})} / \sum_{j} freq{(u,l_{j})}
\end{equation}

To obtain the similarity between two user's profiles $A$ and $B$ and calculate the similarity between two vectors, we consider the cosine similarity method as follows:

\begin{equation}
  cos(\theta)=\frac{A \cdot B}{||A|| ||B||}
  \label{eq:cosine}
\end{equation}

The more similar the two vectors are, the cosine value is closer to $1$, and the less similar they are, its value is closer to $0$. We consider the user's profiles and visited locations as two vectors to calculate the similarity of profiles among users using the cosine similarity method. By examining the similarity of user profiles on the Yelp dataset used in our experiments, we found that the average similarity of users with their friends is 10\%, and in the Gowalla dataset, there is an average of 12\% similarity.

\begin{figure}[H]
  \centering
  \includegraphics[scale=.6]{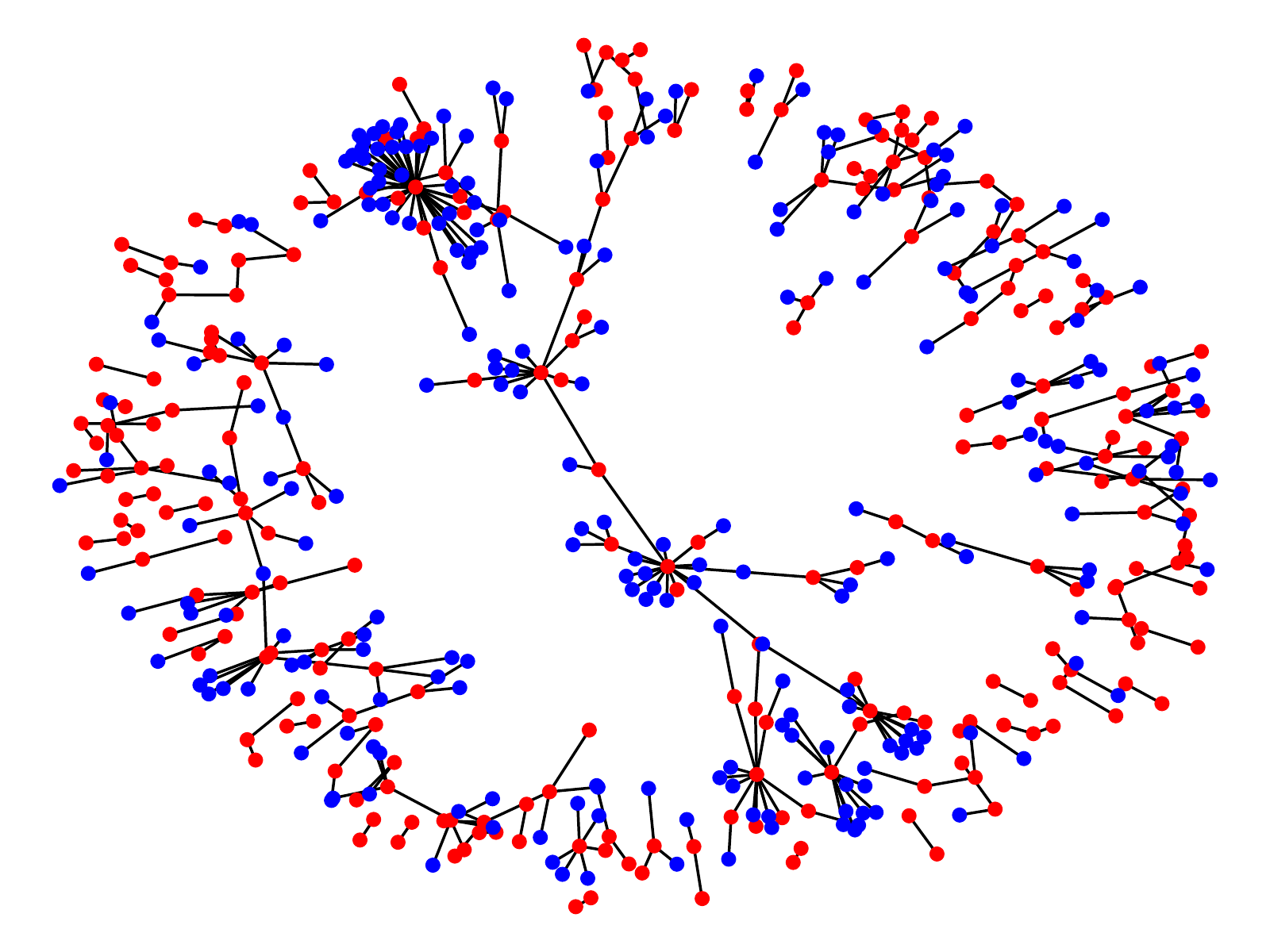}
  \caption{The users and location graph shows the relation between the users, POIs, and visited locations by users and their friends.}
  \label{fig:graph}
\end{figure}

Figure \ref{fig:graph} shows the investigation on the graph of the relationships between users and their visited locations on the Yelp dataset. The results show there are similarities between the locations visited by friends and friends who visit similar locations.

\begin{figure}[H]
\centering
\includegraphics[scale=.6]{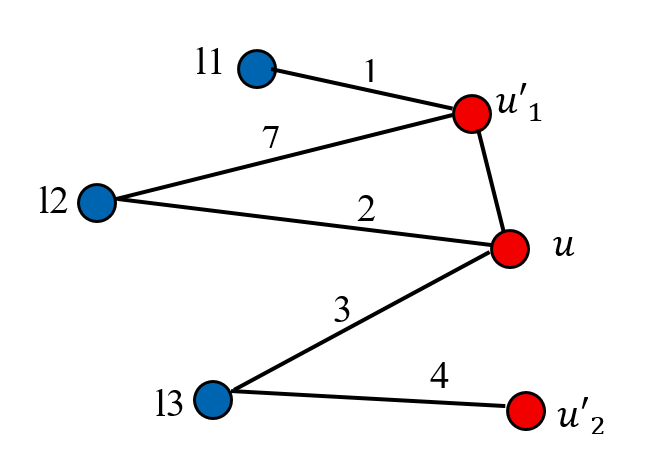}
\caption{Original graph G. Red nodes show users while the blue nodes show locations. An edge between location and user indicates the visited location by that user. Also, an edge between two users shows the friendship in which this edge is unweighed.}
\label{fig:G}
\end{figure}

As we see in Figure \ref{fig:graph}, there are $150$ random users which are shown by red nodes, and about $250$ locations by blue nodes where locations are visited by users and their friends. Also, in Figure \ref{fig:G}, we show a sample relationship between users and visited locations extracted from Figure \ref{fig:graph}, in the case of friendship between two users, an edge connects them, and if the user visits one of the locations, an edge must be drawn between that user's node and the POI.

Inspired by \citet{wang2013location}, we use the integration of similarity and friendliness of users to obtain the effects of social information between users in choosing the locations they would visit. To obtain the friendship influences, in addition to the friendship edge between users, we draw a weighted edge that shows the similarity of their choices. We compute each similarity edge with a tentative weight which is given as follows:

\begin {equation}
     sim (u, u^{\prime}) = cos(pr_{u}, pr_{u^{\prime}})
     \label{eq:sim}
\end {equation}

\begin{figure}[H]
\centering
\includegraphics[scale=.6]{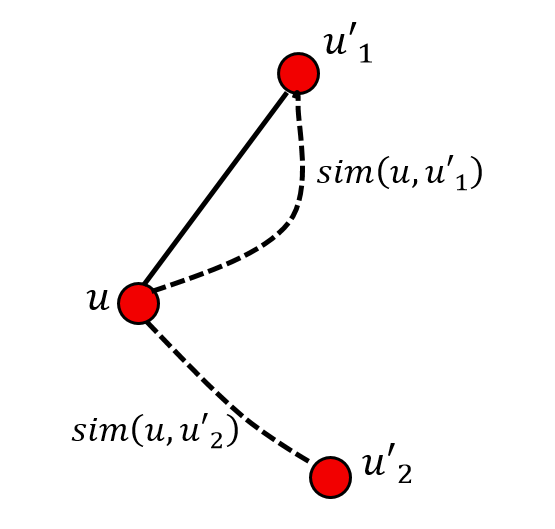}
\caption{Weighted graph of similarity and friendship between users}
\label{fig:graph_social}
\end{figure}

In this regard, $pr_{u}$ and $pr_ {u^{\prime}}$ represent the profile vector of two users which is shown in Eq.~\ref{eq:PSocial}. In this case, there may be both a friendship edge and a similar edge between two users. This results in a graph like Figure~\ref{fig:graph_social}.

Next, we fuse the edge of similarity and the edge of friendship between users. We introduce a parameter $\beta$ to balance the contribution of the implicit and explicit friendship edges. More specifically, for a user $u$, the implicit friend edges are considered with a weight of $\beta$, whereas the explicit friendship links take the weight of $1 - \beta$. We tune this hyperparameter for each dataset on the validation set.
$F_{u}$ is a collection of user $u$'s friends, $LF_{u}$ is the collection of friends who visited locations shared with the user $u$, and  $S_{u}=\sum_{u^{\prime} \in LF_{u}} \operatorname{sim}(u, u^{\prime})$, then the probability of our target user visiting the desired locations will be as follows:

\begin{equation}
p_{u, u '  }=
\begin{cases}
(1-\beta) \cdot \frac{1}{|F_{u}|}, \quad if \; u ' \in \frac{F_{u}}{LF_{u}} 
\\
(\frac{(1-\beta)}{|F_{u}|} + \frac{\beta}{S_{u}} \cdot sim(u,u ')), \quad if \;  u ' \in {F_{u}}\cap{LF_{u}}
\\
\frac{\beta}{S_{u}} \cdot sim(u,u '), \quad if \; u ' \in \frac{LF_{u}}{F_{u}}
\end{cases}
\label{eq:similarity_and_frienship}
\end{equation}

\begin{figure}[H]
\centering
\includegraphics[scale=.6]{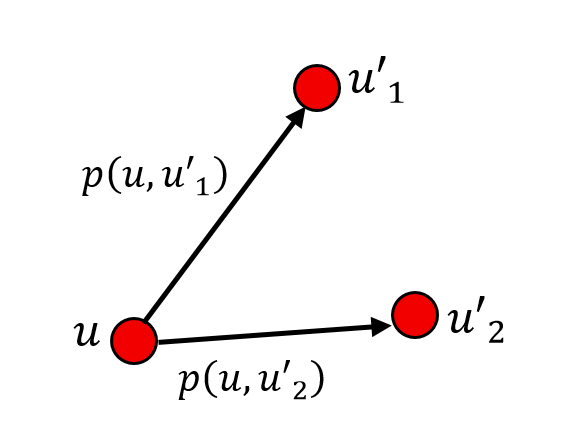}
\caption{Graph $\hat{G}$ is the final graph after combining the friendship edge and the similarity edge between the user and her friends. In this graph, all edges between users are weighed.}
\label{fig:Ghat}
\end{figure}

To this end, we generate a new graph called $\hat{G}$ for each user to show the probability of like-minded between nodes $u'_{1}$ and $u'_{2}$ with node $u$ in the graph.
We indicate the explicit and implicit friendship links as graph edge weights $p$, as shown in Figure~\ref{fig:Ghat}.

Moreover, users' behaviors show that in addition to being influenced by the opinions of their friends, they usually check in around several centers which shows their visited centers depend on the different temporal states. Inspired by \citet{rahmani2020joint}, we make use of two MF models named static and dynamic MF. In the static MF model, the same traditional model \cite{koren2008factorization} is considered for modeling the static user behavior. As mentioned earlier, the purpose of the MF method is to find two matrices with smaller dimensions $U \in \mathbb{R}^{K\times|\mathcal{U}|}$ and $L \in \mathbb{R}^{K\times|\mathcal{L}|}$ based on the user-POI check-ins matrix $R$ such that $R \approx U^{T}L$. Accordingly, the probability that a user $u$ would be interested in a POI $l$ is obtained by $U_u^TL_l$. To model dynamic (or time-depend) user's behavior, user-POI check-ins matrix $R$ divided into sub-matrices $H_1, ..., H_t$ based on the user's check-ins in the different temporal state $t$ such as working or leisure time, weekday or weekend etc. Then, we apply traditional MF on each sub-matrix $H_t$. Finally, it sums all the obtained values into a matrix $\hat{R}$ using the aggregation function, which can be the sum of the resulting matrices. This final matrix $\hat{R}$ indicates the user's preferences on different POIs. Finally, to achieve the user's static and dynamic behavior based on their check-in frequency we fuse the static and dynamic resulting matrices using a product rule as follows:

\begin{equation}
    Z_{u,l} = \hat{R}_{u,l} \times U_u^TL_l
    \label{eq:MF_compute}
\end{equation}

Then, we fusion the proposed contextual information and improve the performance of the POI recommendation system. Algorithm \ref{alg:model} summarizes the basic steps of \modelname. We transmit the main graph $G$ to the $\hat{G}$ graph, which is generated based on the POI visited by the user and their friends. To this end, an element $\pi\textsubscript{j}$ that reflects how node $j$ is close to node $i$ is calculated using Personalize Page Rank (PPR) method \cite{wang2013location} (line $3$). Then, we merge the similarity and friendship edges, and we achieve the effectiveness of social relations between users (line $8$), which along with the influence of activity centers of users depends on the time (line $17$). Finally, we compute the user $ u$'s preference on POI $l$ (line $19$).

\begin{algorithm}
	\DontPrintSemicolon
	\SetAlgoLined
	\SetKwInOut{Input}{Input}\SetKwInOut{Output}{Output}
	\Input{\textit{U, L, G, L$_u$, d, $\beta$}}
	\Output{user-POI preference matrix {$SUCP$}}
	\tcc{$d$ is a threshold for geographical distance}
	Initialize social model $S$ \;
	Normalized checkins matrix as $NM$ \;
   Generate graph $\hat{G}$ with parameter $\beta$ and PPR method \;
   Initialize s\textsubscript{l}=0  for each  $l\in{L}/{L\textsubscript{u}}$ \;
    \ForEach{$u\in{U}$}{
   \ForEach{${u^{\prime}} \in{U}/{\{u\}}$}{
       s\textsubscript{l} = p($u$,{$u'$}) (Eq.~\ref{eq:similarity_and_frienship}) \;
	    \ForEach{$l\in{L\textsubscript{$u' $}}$}{
		    $s\textsubscript{l}= s\textsubscript{l} + \pi\textsubscript{$u'$}. w\textsubscript{$u'$,l}$ \;
    }
    }     
    $ S\textsubscript{u,l}$= $ NM $ .$s\textsubscript{l}$ \;

    	Rank all check-in locations in $|L\textsubscript{u}|$ according to visiting frequency then select user's most frequently check-in as FL\textsubscript{u} \;
		\ForEach{$l\in{L}/{L\textsubscript{u}}$}{				\If{distance($l$, FL\textsubscript{u}) $<d$}{	
                That region becomes center as C\textsubscript{u} in time t
                
                $TC\textsubscript{u,l}$= SumAll(freq\textsubscript C\textsubscript{u,t}/SumAllFreq(C\textsubscript{u,t})(1/dist(l, C\textsubscript{u,t})))
                \;
                        
                $Z\textsubscript{u,l}$ = static and temporal user's preference model (Eq.~\ref{eq:MF_compute})
                
                $SUCP_{u,l}$ = 
                $S\textsubscript{u,l} \cdot TC\textsubscript{u,l} \cdot Z\textsubscript{u,l}$     
                (Eq.~\ref{eq:fusion})
    	}
 }}

 \Return $SUCP$

\caption{\modelname Algorithm}
\label{alg:model}
\end{algorithm}

We select the center that had the most visiting activity among several user activity centers then we apply the same for the user friends. Finally, we obtained a geographical map of the user's activity centers and their friends. Next, we analyze the similarity between users and their friends as well as the distance of their centers. According to the studies as shown in Figure \ref{fig:close_dist}, the higher the level of users and their friends' similarity obtained based on the cosine similarity, the closer the two user's centers are. The proximity of user activity centers to each other and the fact that there are more common locations between friends -- that shows  friendship and similarity -- are directly related to each other, and friends who have activity centers close to each other have more common interests.

\section{Experimental Setup}
\label{sec:evaluation}
In this section, we first present the datasets that we use to compare the methods. Then, we describe the baseline methods and finally, we introduce evaluation metrics.

\subsection{Datasets}
We conduct experiments using two publicly available real-world LBSN check-in datasets from Yelp and Gowalla provided by the authors of~\cite{liu2017experimental}\footnote{\url{http://spatialkeyword.sce.ntu.edu.sg/eval-vldb17/}}.

\paragraph{Gowalla dataset}
The Gowalla dataset includes user's check-ins information from February 2009 to October 2010. Following \cite{liu2017experimental,manotumruksa2018contextual}, we consider the threshold to preprocess the dataset for users and POIs to visit by removing users who visited locations for less than $15$ check-ins because they have very little activity. We exclude POIs of less than $10$ which may cause a spam error on the model. Therefore, the Gowalla dataset, finally, includes 31,803 POIs, 5,628 users, 620,683 check-ins, 46,001 social links and the sparsity level is 99.78\%.

\paragraph{Yelp dataset}
We find that the Yelp dataset is well-defined and includes all the studied contextual information (the geographical coordinates, POI category, friendship information, and the check-in timestamp). The Yelp dataset is provided by the Yelp dataset challenge\footnote{\url{https://www.yelp.com/dataset/challenge}} round 7 (access date: Feb 2016) in 10 metropolitan areas across two countries. The threshold that we intended for the preprocessing of the Yelp dataset includes the removal of users who have less than $10$ visited locations and for POIs similar to the Gowalla dataset, we removed with less than $10$ visits. After this preprocessing, the Yelp dataset includes 7,135 users, 16,621 POIs, 301,753 check-ins, and 46,778 social relations. The sparsity of the user-POI check-in matrix is 99.86\%.

\subsection{Social links data leakage} \label{sec:overlap}
In this section, we discuss the possibility of data leakage when considering the explicit friendship available in both datasets. The available social information is not accompanied by temporal information. Therefore, while it is explicitly indicated which users are friends, the information about when these users followed each other on the platform is missing. Therefore, it is possible that two users follow each other during the test period of the dataset, but their friendship information is used in the training phase. 

We aim to analyze and characterize the extent of data leakage probability by considering the percentage of check-in overlap that two users have in the training data. Inspired by \citet{ji2020critical}, we take into consideration the number of check-ins two users have in common during the training period and consider them as \emph{training friends} only if their training overlap exceeds a certain threshold. To do this, we sort the check-ins of each user in chronological order and compare two users based on the earliest training period. For example, assume the training check-ins of user $u_1$ is in the period of 2011-01-01 till 2011-02-15, and the training check-ins of user $u_2$ occur between 2011-01-10 and 2011-04-18. In this example, we take 2011-02-15 as the end of the training period of these two users (because it is the earliest of the two) and count the number of check-ins dated before 2011-02-15. We then take the number of the training check-ins of both users and divide it by the total number of their check-ins to compute the percentage of \emph{training check-in overlap} of the two users.

\begin{figure}[!tbp]
  \centering
  \subfloat[Overlap on the number of friendships in Gowalla]{\includegraphics[width=0.46\textwidth]{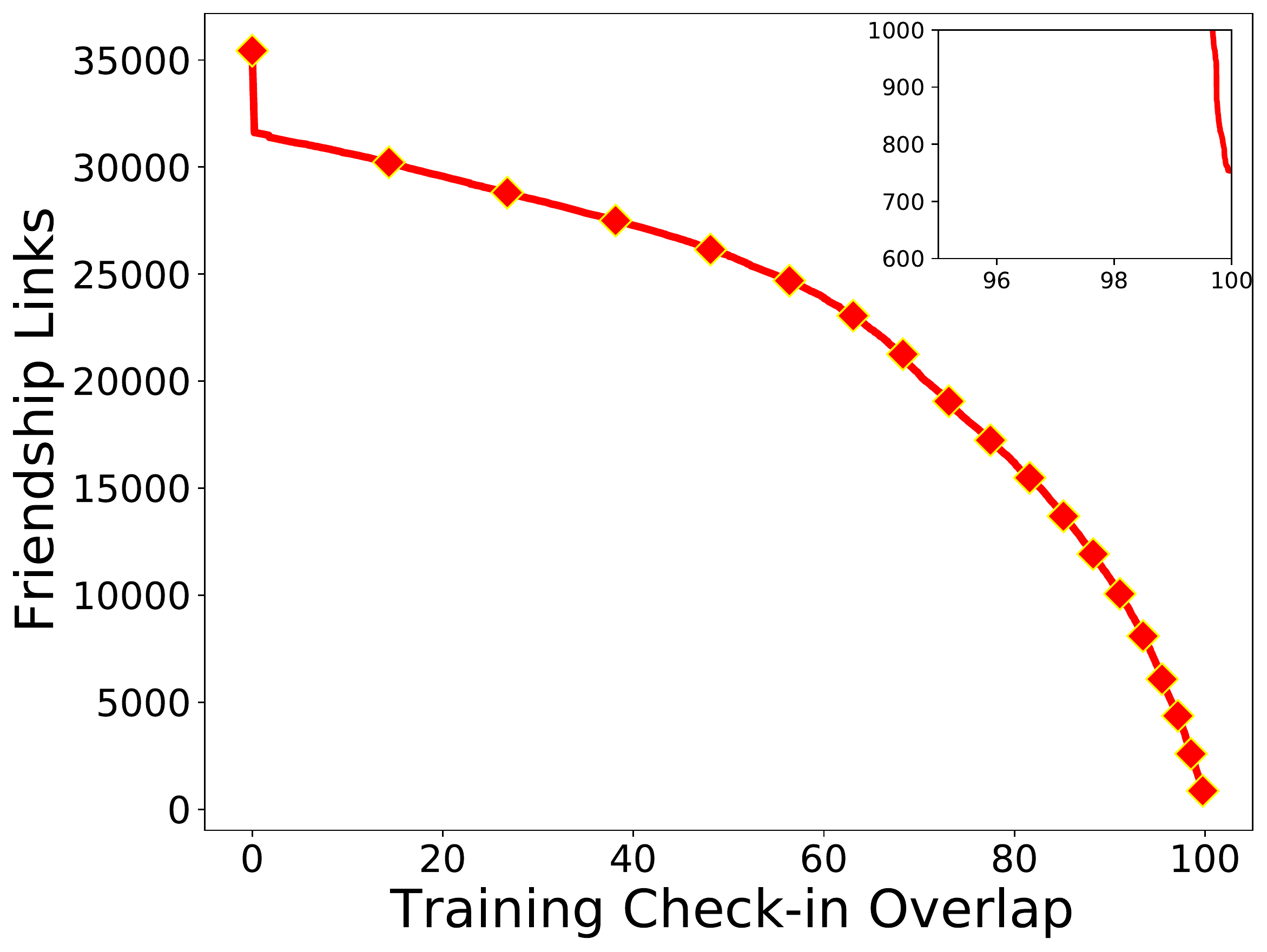}\label{fig:OF1}}
  \hfill
  \subfloat[Overlap on the number of friendships in Yelp]{\includegraphics[width=0.46\textwidth]{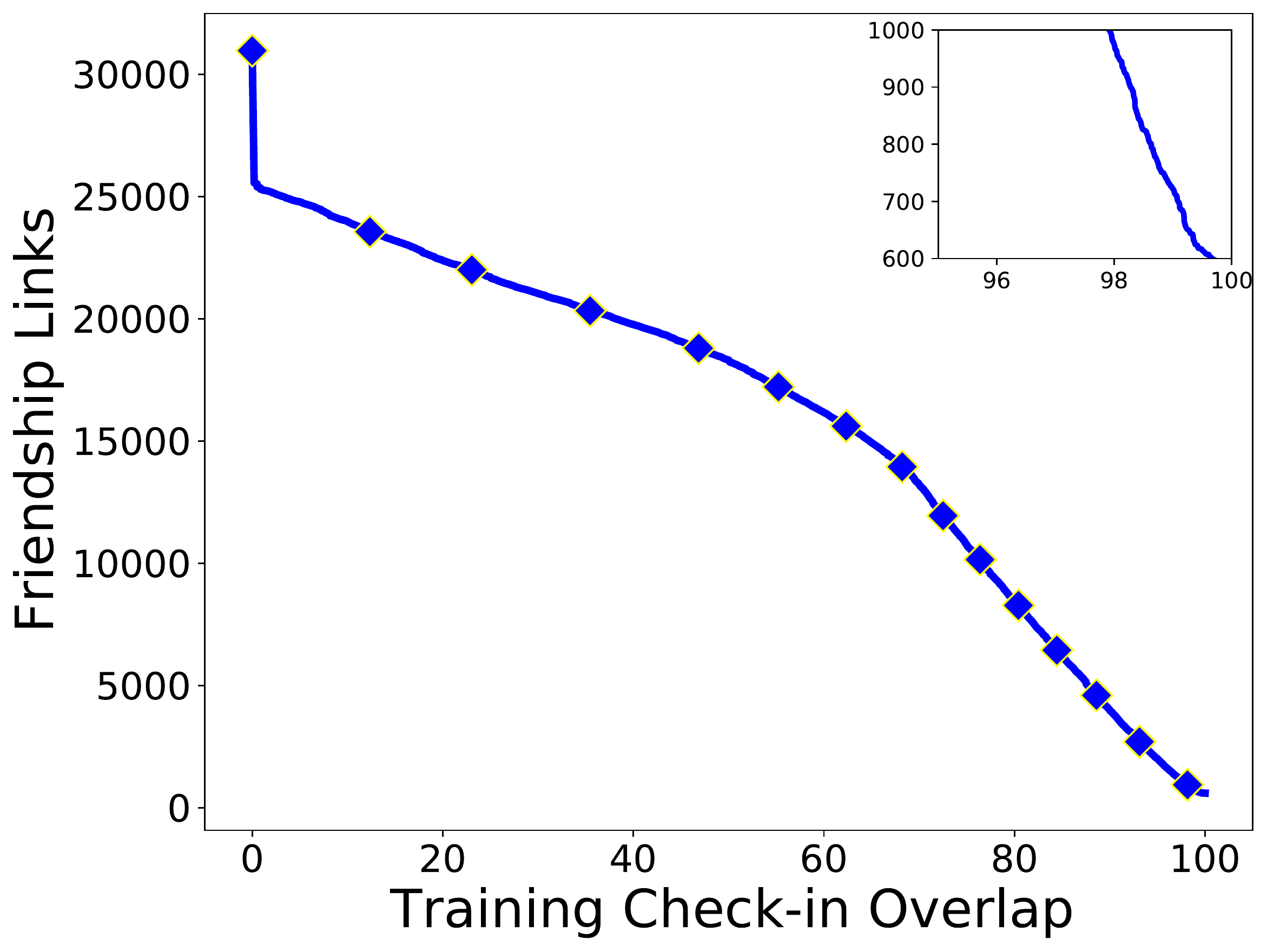}\label{fig:OF2}}
  \caption{POIs overlap on the number of friendships.}
  \label{fig:overlap Frienship}
\end{figure}

Figure~\ref{fig:overlap Frienship} plots the number of validated training friendship links for different values of overlap threshold. We can see that in this plot if the threshold is zero, we have all the available friendship links. However, we see an immediate drop as we take a threshold of $\ge 0$, indicating that a considerable portion of users did not have any overlap in the training period, but still been friends. Such cases happen when user $u_1$ is active for a while on the platform but then leaves the platform (no check-in data available). After that user leaves the platform user $u_2$ joins the platform and follows that user ($u_1$ also follows $u_2$ but perhaps during their testing period). Therefore, it is very likely that $u_1$ follows $u_2$ during their testing period, hence leading to data leakage. Indeed such a case is when we can claim with high confidence that such a link leads to data leakage. However, we cannot still be sure about other cases with more check-in overlap. Even though these users are more likely to have followed each other during the training period, we cannot claim that it has happened during the testing period. Nevertheless, as we see in the plot, for both datasets, as the overlap threshold increases, the less valid explicit friendship links remain. In this work, we take a rather strict threshold of 70\% check-in overlap between two users to keep their friendship link in the training data. This is to minimize the chance of data leakage. Note that, as mentioned earlier we cannot guarantee that it does not happen as this requires timestamped friendship information. We see in Section~\ref{sec:leakage} how this experiment affects the performance of our model compared to the social baseline and the possibility of data leakage based on those results.

\subsection{Baselines}
We compared the proposed \modelname with advanced POI recommended approaches that consider social or the other contextual effects in the recommendation process. We also compare our proposed \modelname with MF-based methods, Multi-Layer Perceptron (MLP) neural network, and Recurrent Neural Network (RNN) based approaches. Table \ref{tbl:baselines} characterizes the various baselines into different categories, namely neural network-based approaches, context-aware-based approaches, and indicates the contextual information of the models. We summarized the details of the baselines as follows:

\begin{itemize}
    \item \textbf{TopPopular} \cite{dacrema2019we}: A non-personalized method that recommends the POIs that have the highest number of visits.
    
    \item \textbf{PFMMGM} \cite{cheng2012fused}: A method that fuses the Passion Factorization Model (PFM) with the geographical information-based model. PFMMGM models the user's geographical behavior around several centers that these centers include most of the user's check-ins.
    
    \item \textbf{LRT} \cite{gao2013exploring}: A method that learns the user's and POIs' factors based on temporal information and incorporates temporal influence in a latent ranking model.
    
    \item \textbf{LMFT} \cite{stepan2016incorporating}: A method that gives a higher rank to the POIs that have recent visits and multiple check-in frequencies.
    
    \item \textbf{GeoSoCa}\footnote{To evaluate GeoSoCa on the Gowalla datasets we remove the categorical model of GeoSoCa (i.e., \textit{Ca}) as the Gowalla dataset does not include the categorical data.} \cite{zhang2015geosoca}: A method that includes three contextual information, i.e., geographical, social, and categorical. GeoSoCa models the geographical influence as a kernel estimation method on each user incorporates the check-in frequency of a user's friends as a social influence and fuses categorical information to gain POI popularity.
    
    \item \textbf{iGLSR} \cite{zhang2013igslr}: A method that applies kernel density estimation based on social and geographical influence on POI recommendation.
    
    \item \textbf{Rank-GeoFM} \cite{li2015rank}: A method that ranks recommended POIs based on the geographical information and MF model that includes the geographical influence of neighboring POIs.
    
    \item \textbf{MLP} \cite{he2017neural}: A method that uses an MLP neural network instead of dot produce to add the non-linearity to user-POI interaction.
    
    \item \textbf{L-WMF} \cite{guo2019location}: A method that incorporates the geographical neighborhoods into the WMF model as a regularization term. This can add geographical information from a POI perspective to the model. 
    
    \item \textbf{LFBCA} \cite{wang2013location}: A method that creates a graph that adds an edge between users and their friendship, as well as users and locations. The constructed graph includes different types of edges to show both user preference and social influence.
    
    \item \textbf{CARA} \cite{manotumruksa2018contextual}: An RNN-based method considers both sequences of check-ins and contextual information associated with the sequences to capture the user's dynamic preferences.
    
    \item \textbf{\modelname-NoSocial}: A variant of our \modelname that models geographical and temporal information, but does not include social information.
\end{itemize}

\begin{table}[]
\caption{Summary of baselines}
\label{tbl:baselines}
\begin{adjustbox}{max width=\textwidth}
\begin{tabular}{lcccccc}
\toprule
\multirow{2}{*}{\textbf{Baselines}}                     & \multirow{2}{*}{\textbf{Neural Networks}} & \multirow{2}{*}{\textbf{Context-aware}} & \multicolumn{4}{c}{\textbf{Context Information}}        \\ \cline{4-7} 
                                               &                                  &                                & \textit{Geographical} & \textit{Temporal} & \textit{Social} & \textit{Categorical} \\
\midrule
\textbf{TopPopular} \cite{dacrema2019we}       &$\times$                          &$\times$                        & $\times$     & $\times$ &$\times$& $\times$     \\
\textbf{PFMMGM} \cite{cheng2012fused}            &$\times$                          &$\checkmark$               & $\checkmark$ & $\times$ &$\times$& $\times$     \\
\textbf{LRT} \cite{gao2013exploring}           &$\times$                          &$\checkmark$               & $\times$     &$\checkmark$&$\times$& $\times$   \\
\textbf{LMFT} \cite{stepan2016incorporating}   &$\times$                          &$\checkmark$               & $\times$     &$\checkmark$ & $\times$ &$\times$       \\
\textbf{GeoSoCa} \cite{zhang2015geosoca}       &$\times$                          &$\checkmark$               & $\checkmark$ &$\times$&$\checkmark$&$\checkmark$\\
\textbf{iGLSR} \cite{zhang2013igslr}           &$\times$                          &$\checkmark$              & $\checkmark$ & $\times$         & $\checkmark$       &$\times$     \\
\textbf{Rank-GeoFM} \cite{li2015rank}          &$\times$                          &$\checkmark$              & $\checkmark$ & $\times$&$\times$  &$\times$              \\
\textbf{MLP} \cite{he2017neural}               &$\checkmark$                 &$\times$                        & $\times$     &$\times$&$\times$&$\times$        \\
\textbf{L-WMF} \cite{guo2019location}          &$\times$                          &$\checkmark$              & $\checkmark$ & $\times$ & $\times$&$\times$      \\
\textbf{LFBCA} \cite{wang2013location}         &$\times$                          &$\checkmark$              & $\times$     & $\times$&$\checkmark$ & $\times$ \\
\textbf{CARA} \cite{manotumruksa2018contextual}&$\checkmark$                 &$\checkmark$              & $\times$     & $\checkmark$& $\times$ &$\times$      \\
\bottomrule
\end{tabular}
\end{adjustbox}
\end{table}

\subsection{Evaluation Metrics}
\label{sec:evalmetrics}
We evaluate the performance and accuracy of our proposed model and the baseline methods using three common evaluation metrics, namely, \textit{Precision at N} (Precision@N), \textit{Recall at N} (Recall@N), and \textit{Normalized Discounted Cumulative Gain at N} (nDCG@N) with $N \in \{10, 20\}$ - as applied in previous literature \cite{rahmani2020joint,liu2017experimental,aliannejadi2017personalized}. The precision is a measure of recommendation relevancy, while recall is a measure of how many truly relevant recommendations are returned. The measure nDCG@N evaluates the ranking quality of the recommendation models. 
We follow the common practice from previous works \cite{rahmani2019lglmf,liu2017experimental,sanchez2021point} to evaluate the proposed model and baselines, we split the Gowalla and Yelp datasets into three different parts for each user, i.e., \textit{train}, \textit{valid}, and \textit{test} sets. To this end, we sort the check-ins of each user in chronological order and take the 20\% most recent check-ins as the test set, the next 10\% as the validation set, and the rest as the training set.
We also use the paired t-test $(p < 0.05)$ to indicate the statistically significant differences in the results. Suppose we return a recommendation list that includes top-N recommended POIs for user $u$, Precision@N is defined as: 

\begin{equation}
    Precision@N = \frac{tp_u}{tp_u + fp_u}
\end{equation}
\noindent
and Recall@N is defiend as:

\begin{equation}
    Recall@N = \frac{tp_u}{tp_u + tn_u}
\end{equation}
\noindent
in which $tp_u$ is the number of recommended POIs that are visited by $u$ and $fp_u$ is the number of recommended POIs that are not visited by $u$. $tn_u$ is also the number of POIs visited by $u$ but not in the top-N recommendations. In this paper, we report the average precision and recall values of all users.

Also, for each user, nDCG@N is defined as: 

\begin{equation}
    nDCG@N = \frac{DCG@N}{IDCG@N}
\end{equation}
\noindent
where $DCG@N=\sum_{i=1}^N\frac{2^{rel_i}-1}{log_2(i+1)}$ and IDCG@N is equal to the value of ideally ranked DCG@N. Also, $rel_i$ indicates the relevancy of the item that is ranked at the position $i$. The higher value (in the range $0$ to $1$) of nDCG@N means better results. The average nDCG values of all users are reported.

\begin{table*}[!htb]
\settowidth\rotheadsize{FouesquareDataset}
  \caption{The performance comparison of \modelname and its variant with baselines based on Precision@$N$, Recall@$N$, and nDCG@$N$ for $N \in \{10,20\}$ on the Gowalla and Yelp datasets. The superscripts $\dagger$ and $\ddagger$ show the significant improvements compared to baselines and model variation, respectively ($p < 0.05$). The reported results for Gowalla are based on $\beta =0.7 $ and for Yelp are based on $\beta =0.8 $.}
  \label{tbl:results}
  \begin{adjustbox}{max width=\textwidth}
      \begin{tabular}{ll@{\quad}r@{\quad}rr@{\quad}@{\quad}r@{\quad}rr@{\quad}@{\quad}r@{\quad}rr}
        \toprule
        \multirow{2}{*}{} & \multirow{2}{*}{\textbf{Baselines}} & \multicolumn{2}{c}{\textbf{Precision}} && \multicolumn{2}{c}{\textbf{Recall}} && \multicolumn{2}{c}{\textbf{nDCG}} \\
        \cmidrule{3-4} \cmidrule{6-7} \cmidrule{9-10}
        & & \multicolumn{1}{c}{\textbf{@10}} & \multicolumn{1}{c}{\textbf{@20}} && \multicolumn{1}{c}{\textbf{@10}} & \multicolumn{1}{c}{\textbf{@20}} && \multicolumn{1}{c}{\textbf{@10}} & \multicolumn{1}{c}{\textbf{@20}} \\
        \midrule
        \multirow{8}{*}[0ex]{\rothead{\textbf{Gowalla}}} & TopPopular & 0.0192 & 0.0146 && 0.0176 & 0.0270 && 0.0088 & 0.0079 \\
        & LRT & 0.0249 & 0.0182 && 0.0220 & 0.0321 && 0.0105 & 0.0093 \\
        & PFMMGM & 0.0240 & 0.0207 && 0.0258 & 0.0442 && 0.0140 & 0.0144 \\
        & MLP & 0.0243 & 0.0215 && 0.0237 & 0.0396 && 0.0098 & 0.0127 \\
        & LMFT & 0.0315 & 0.0269 && 0.0303 & 0.0515 && 0.0157 & 0.0150 \\
        & GeoSoCa & 0.0215 & 0.0195 && 0.0253 & 0.0449 && 0.0222 & 0.0206 \\
        & iGLSR & 0.0297 & 0.0242 && 0.0283 & 0.0441 && 0.0153 & 0.0145 \\
        & L-WMF & 0.0341 & 0.0296 && 0.0351 & 0.0582 && 0.0183 & 0.0178 \\
        & Rank-GeoFM & 0.0352 & 0.0297 && 0.0379 & 0.0602 && 0.0187 & 0.0179 \\
        & LFBCA & 0.0453 & 0.0376 && 0.0460 & 0.0742 && 0.0492  & 0.0427  \\
        & CARA & 0.0501 & 0.0410 && 0.0485 & 0.0792 && 0.0531  & 0.0452  \\
        \hdashline
        & \modelname-NoSocial & 0.0381 & 0.0317 && 0.0403 & 0.0648 && 0.0407  & 0.0355  \\
        & \modelname & \textbf{0.0502}$^{\ddagger}$ & \textbf{0.0410}$^{\ddagger}$ && \textbf{0.0520}$^{\dagger\ddagger}$ &
        \textbf{0.0823}$^{\dagger\ddagger}$ && \textbf{0.0548}$^{\ddagger}$ & \textbf{0.0470}$^{\dagger\ddagger}$ \\
      \midrule \midrule
        \multirow{8}{*}[0ex]{\rothead{\textbf{Yelp}}} & TopPopular & 0.0077 & 0.0073 && 0.0095 & 0.0185 && 0.0075 & 0.0073 \\
        & LRT & 0.0083 &0.0081 && 0.0104 &0.0219 && 0.0089 & 0.0085 \\
        & PFMMGM & 0.0162 & 0.0135 && 0.0242 & 0.0402 && 0.0175 & 0.0152 \\
        & MLP & 0.0203 & 0.0174 && 0.0284 & 0.0517 && 0.0195 & 0.0185 \\
        & LMFT & 0.0183 & 0.0163 && 0.0264 & 0.0457 && 0.0194 & 0.0176 \\
        & GeoSoCa & 0.0183 & 0.0150 && 0.0214 & 0.0340 && 0.0195 & 0.0168  \\
        & iGLSR & 0.0235 & 0.0192 && 0.0306 & 0.0529 && 0.0231 & 0.0214 \\
        & L-WMF & 0.0215 & 0.0181 && 0.0295 & 0.0542 && 0.0202 & 0.0194 \\
        & Rank-GeoFM & 0.0231 & 0.0198 && 0.0316 & 0.0587 && 0.0217 & 0.0214 \\
        & CARA & 0.0255 & 0.0219 && 0.0362 & 0.0628 && 0.0271 & 0.0239  \\
        & LFBCA   & 0.0269 & 0.0228 && 0.0408 & 0.0667 && 0.0290 & 0.0255   \\
       \hdashline
        & \modelname-NoSocial & 0.0251 & 0.0217 && 0.0383 & 0.065 && 0.0264 &  0.0236  \\
        & \modelname & \textbf{0.0286}$^{\ddagger}$ & \textbf{0.0245}$^{\ddagger}$ && \textbf{0.0438}$^{\ddagger}$ & \textbf{0.0731}$^{\dagger\ddagger}$ && \textbf{0.0304}$^{\ddagger}$ & \textbf{0.0270}$^{\ddagger}$ \\
      \bottomrule
    \end{tabular}
\end{adjustbox}
\end{table*}

\section{Results}
\label{sec:results}
This section shows the results of experiments performed on two datasets, Yelp and Gowalla.

\subsection{Performance evaluation against baselines}
We present the results of performance accuracy and the statistical significance test of our experiments in Table \ref{tbl:results}. As we see, our proposed \modelname model significantly outperforms the other compared methods based on all three evaluation metrics on both datasets. TopPopular shows that deep learning methods are not always better than non-personalized methods \cite{dacrema2019we}. Therefore, we include this method to compare the results of \modelname with a non-personalized method. \modelname outperforms PFMMGM by a significant margin that is among the basic methods in POI recommendation systems. Comparing with other geographical-based methods, i.e., iGLSR, L-WMF, and Rank-GeoFM, we see that \modelname demonstrates the best performance. 
The results show that \modelname defeats all methods based on social and geographical relationships in terms of all metrics for all different N values on both datasets. 

For example, the result in the Gowalla dataset with the Recall@10 metric for LRT and iGLSR are respectively 0.0220 and 0.0283, while the result of \modelname is 0.0520, demonstrating the effectiveness of our social model. More importantly, \modelname shows a significant improvement compared to neural network-based models. \modelname achieves better performance than MLP and CARA. \modelname outperforms CARA which has a recurrent neural network-based architecture. This implies that considering more contextual information in POI recommendation can improve the performance of the model.

\subsection{Effect of check-in overlap} \label{sec:leakage}

\begin{figure}[!tbp]
  \centering
  \subfloat[Overlap on nDCG@20 in Gowalla]{\includegraphics[width=0.46\textwidth]{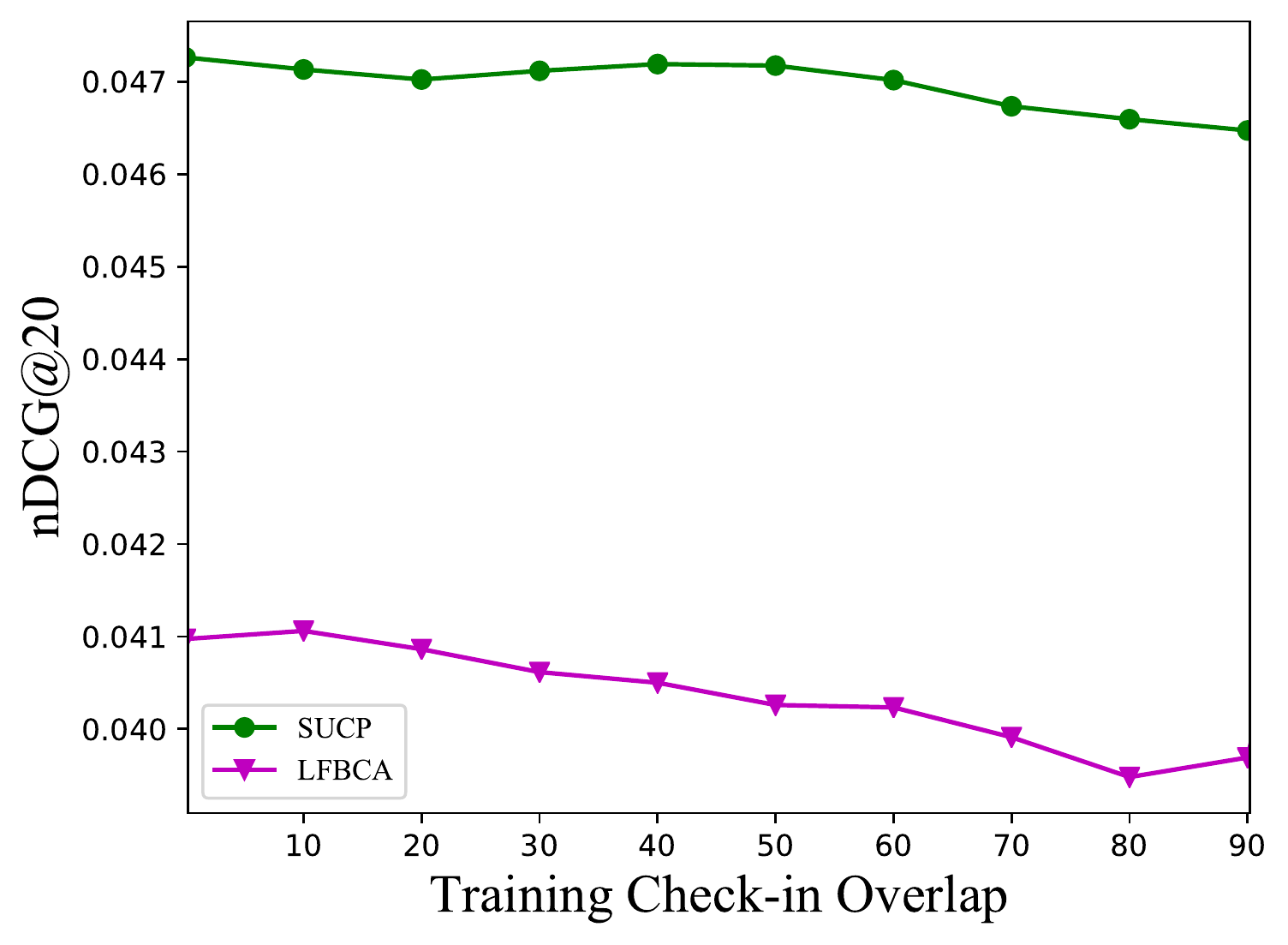}\label{fig:OnD1}}
  \hfill
  \subfloat[Overlap on nDCG@20 in Yelp]{\includegraphics[width=0.46\textwidth]{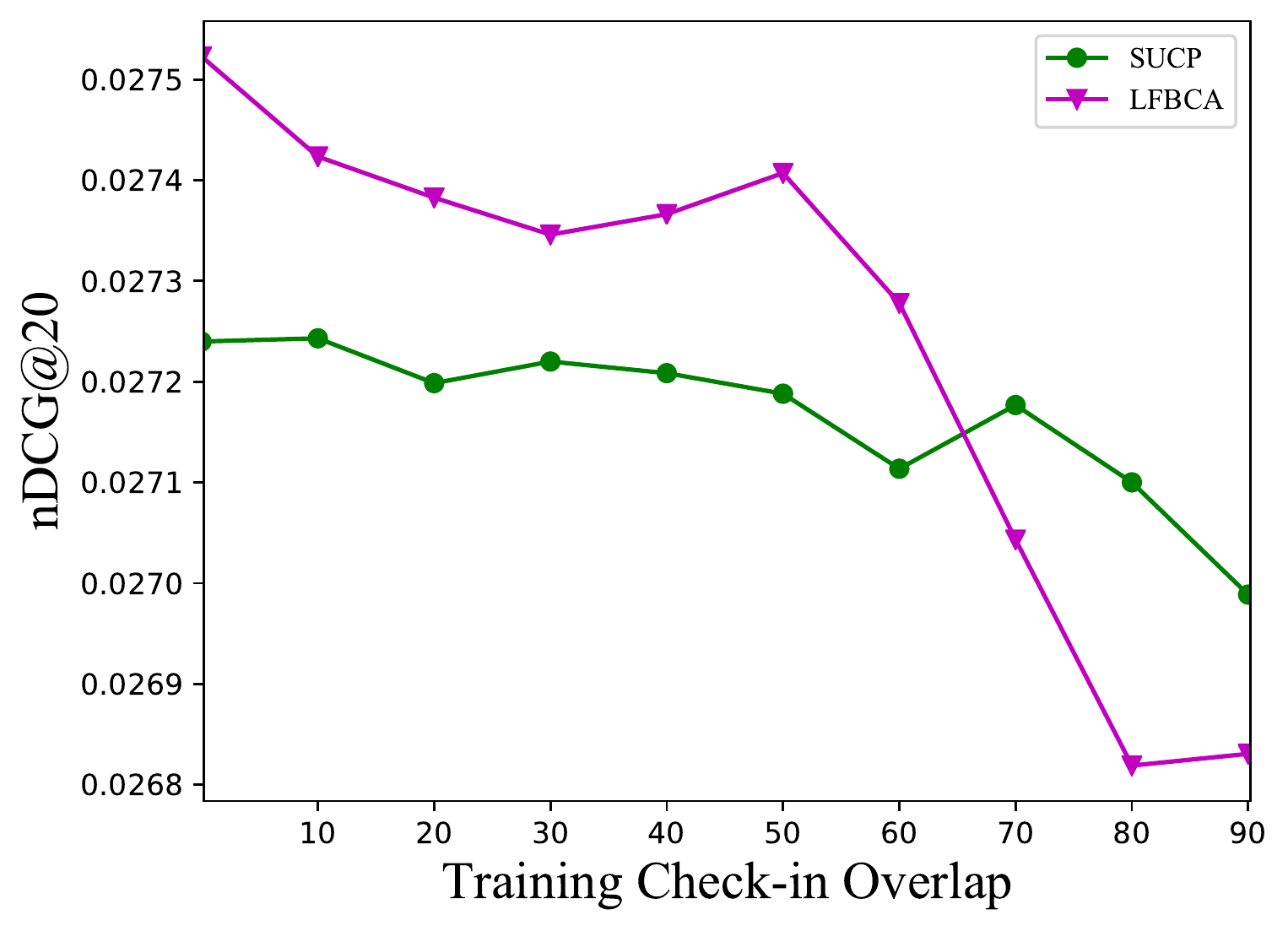}\label{fig:OnD2}}
  \caption{Impact of different overlaps in results of \modelname and LFBCA}
  \label{fig:overlap result}
\end{figure}

In this section, we aim to examine the effect of reducing the number of explicit friendship links in the training data, following our proposed check-in overlap threshold (see Section~\label{sec:overlap}). Here, we aim to compare the performance of our model with the social baseline to compare the sensitivity of both models on the explicit friendship data, and how that affects the performance. 
To do so, we keep all the hyperparameters fixed and change the training data based on the available training friendships for different values of overlap threshold. In Figure~\ref{fig:overlap result}, we see that on both datasets, our \modelname is less affected as the number of explicit friendship links is reduced. The LFBCA model, on the other hand, exhibits a poorer performance as the overlap threshold increases. This suggests that while LFBCA is more dependent on explicit social links and perhaps taking more advantage of possible data leakage, our \modelname which relies on other information is more robust, less affected by the potential data leakage of the explicit social links.

\subsection{Effect of number of visited POIs}
\begin{figure}[!tbp]
  \centering
  \subfloat[Recall@20 on Gowalla]{\includegraphics[width=0.5\textwidth]{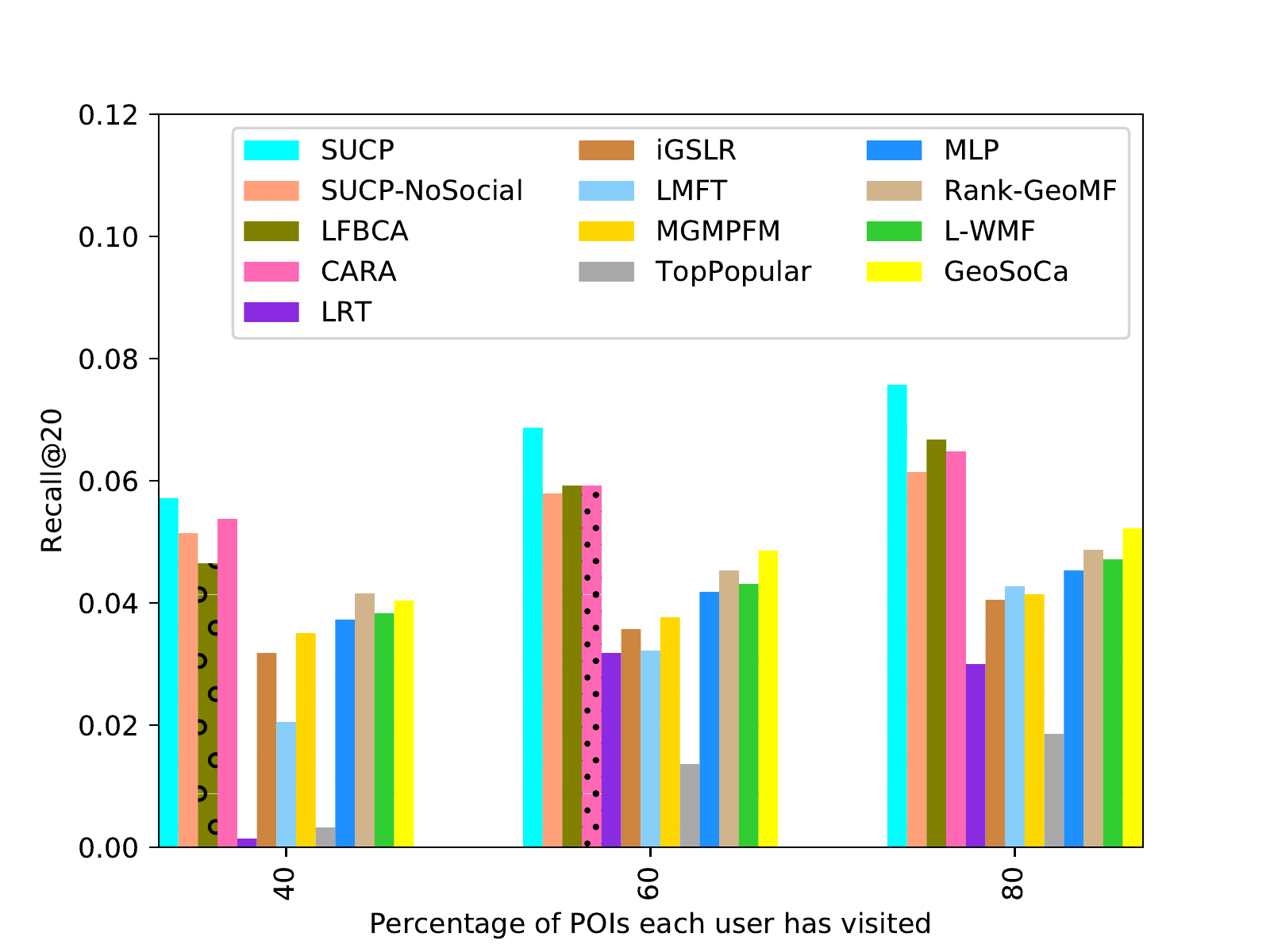}\label{fig:gf1}}
  \hfill
  \subfloat[nDCG@20 on Gowalla]{\includegraphics[width=0.5\textwidth]{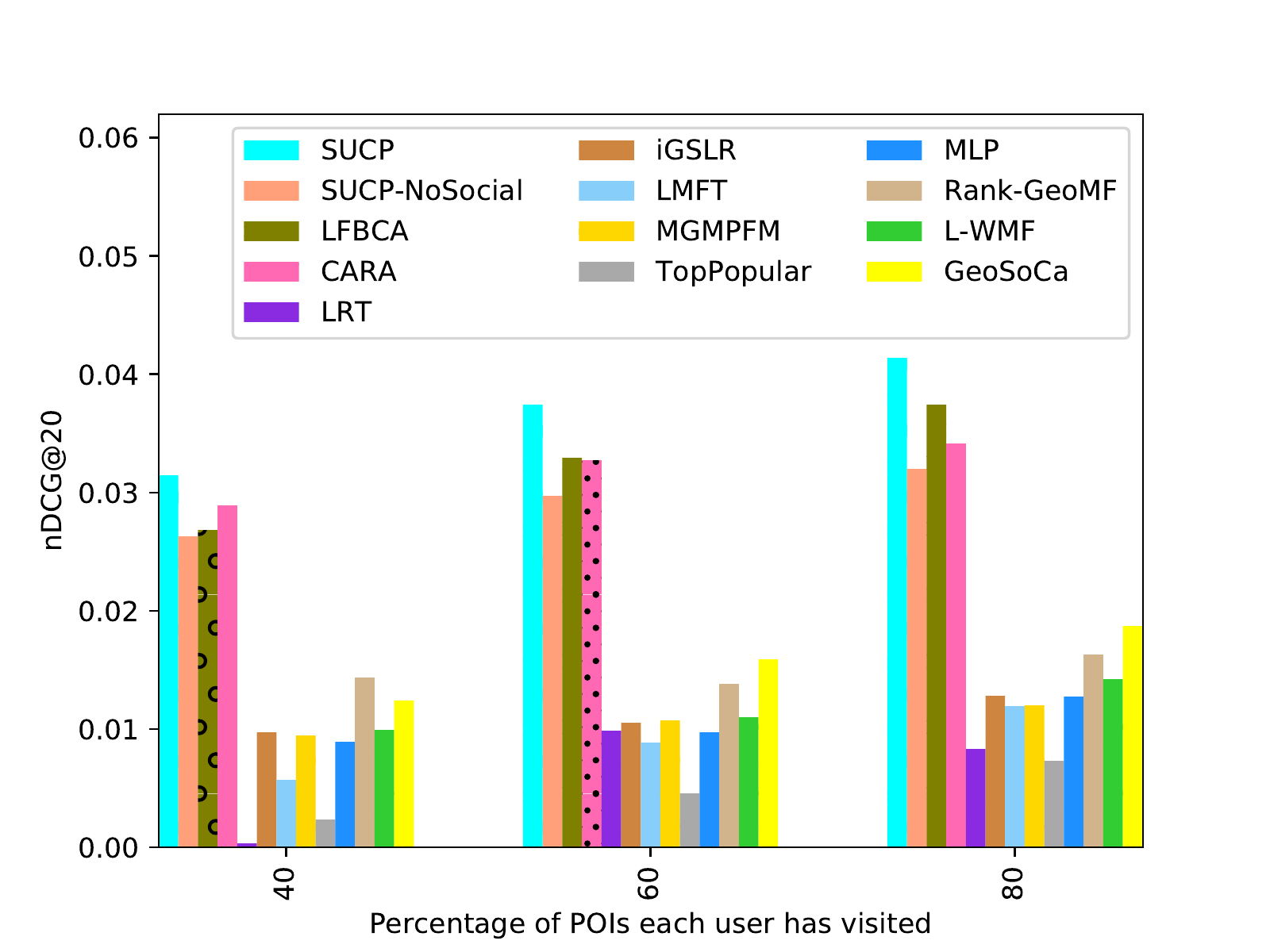}\label{fig:gf2}}
  \caption{Effect of data sparsity on (a) Recall@20 and (b) nDCG@20 with $40\%$, $60\%$, and $80\%$ of POIs that each user has visited on the Gowalla dataset.}
  \label{fig:sparsity-gowalla}
\end{figure}

\begin{figure}[!tbp]
  \centering
  \subfloat[Recall@20 on Yelp]{\includegraphics[width=0.5\textwidth]{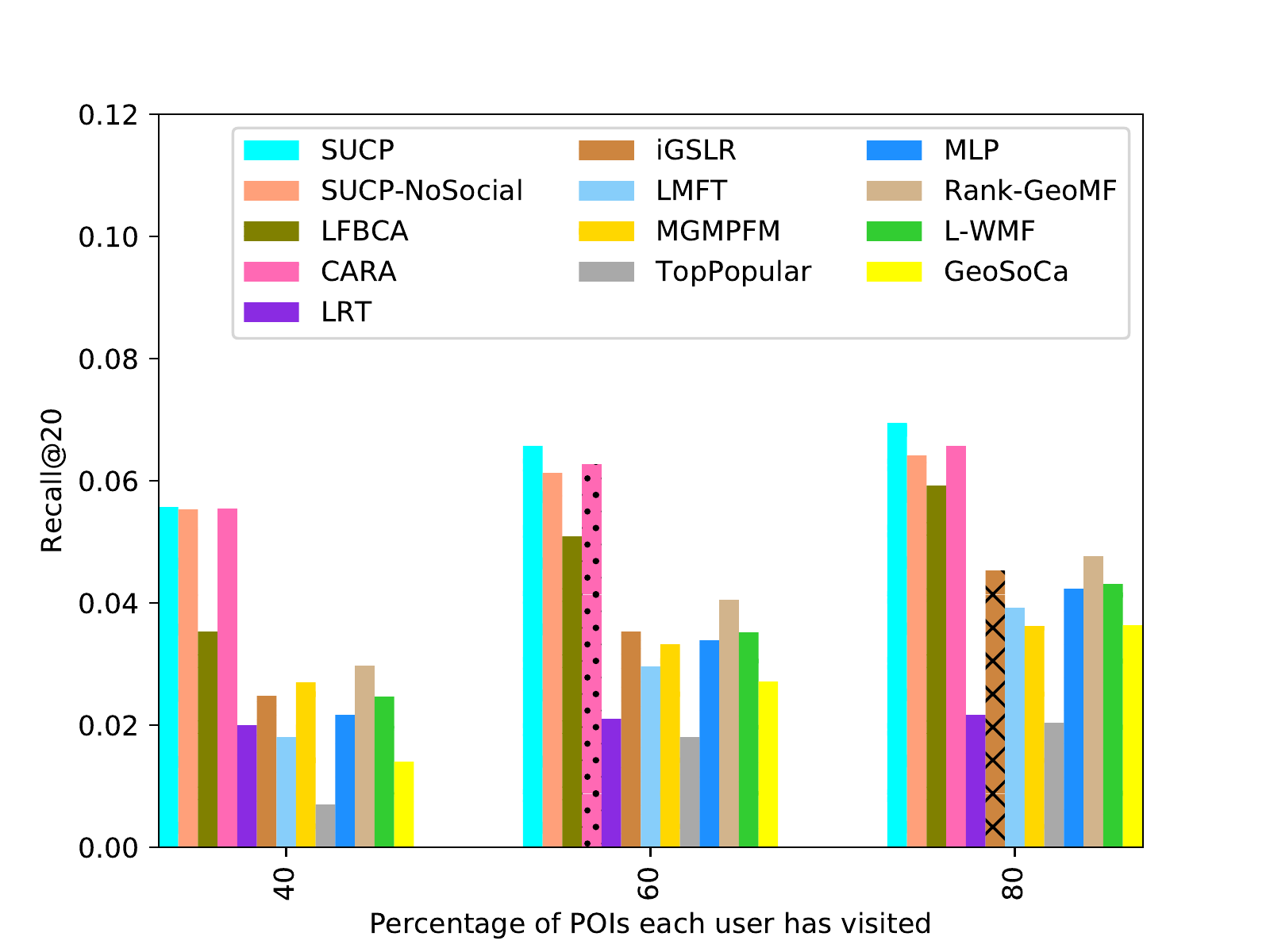}\label{fig:yf1}}
  \hfill
  \subfloat[nDCG@20 on Yelp]{\includegraphics[width=0.5\textwidth]{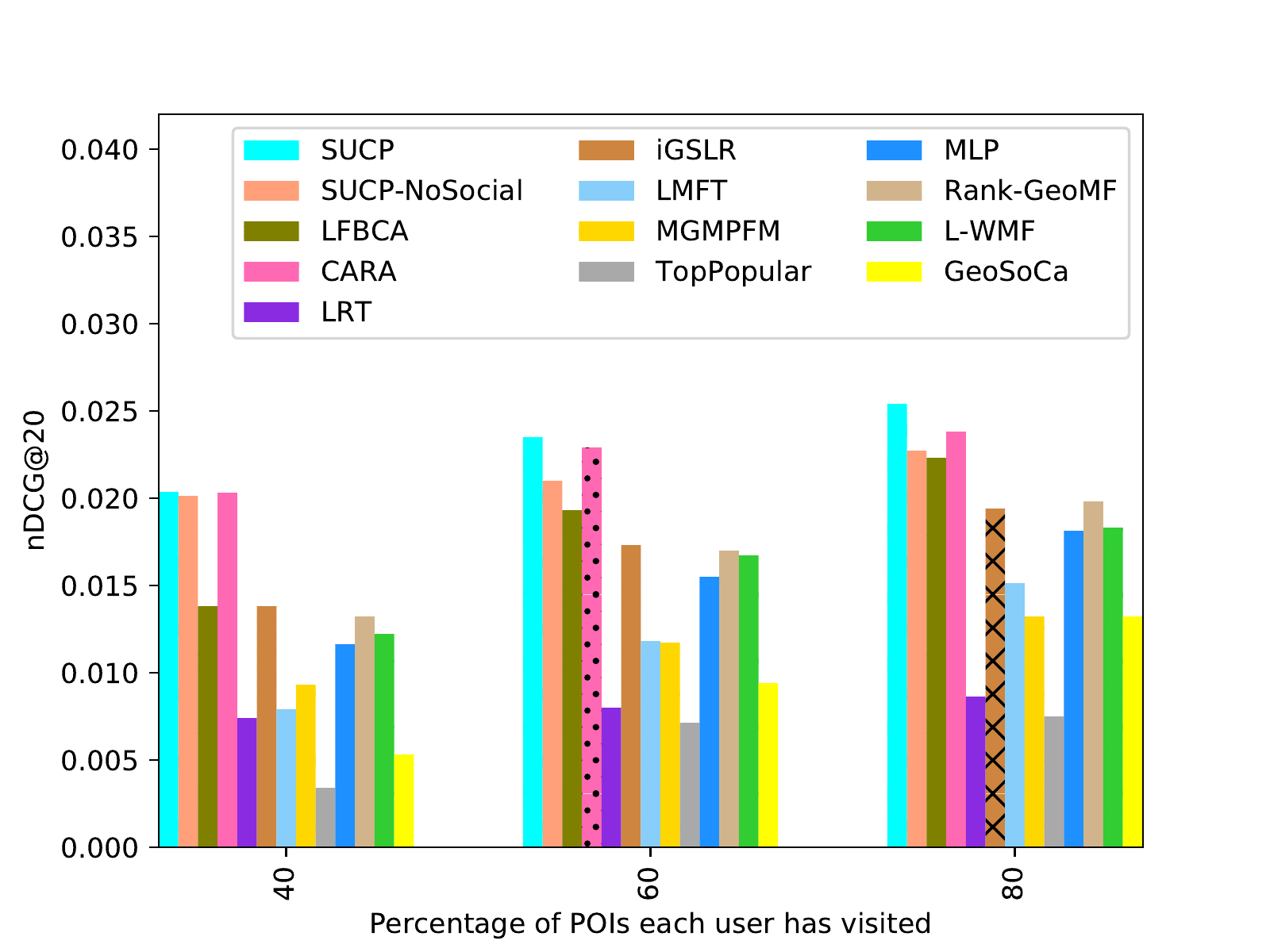}\label{fig:yf2}}
  \caption{Effect of data sparsity on (a) Recall@20 and (b) nDCG@20 with $40\%$, $60\%$, and $80\%$ of POIs that each user has visited on the Yelp dataset.}
  \label{fig:sparcity-yelp}
\end{figure}

We train \modelname and all the baseline methods with different data sizes. To do this, for each user, we consider only a certain percentage of the POIs visited inside the training set at random, from 40\% to 80\%. Figures \ref{fig:sparsity-gowalla} and \ref{fig:sparcity-yelp}report the performance of \modelname and all baseline models on the two datasets. As can be seen in these figures, \modelname performs better than LRT, TopPopular, MLP, and LMFT. As one can see in Figure \ref{fig:gf1}, \modelname is more robust than LMFT on the Gowalla datasets when we consider only $40\%$ of user's check-ins. The accuracy of \modelname and LMFT decrease by about $15\%$ and $26\%$, respectively. Therefore, these results show that the method \modelname in contrast to data sparsity also achieves better results than other baselines and is robust in the case of data sparsity. For example, when trained on 40\% of the data, our \modelname achieves about 300\% improvement over iGLSR.

As mentioned earlier, one of the challenges with recommender systems is data sparsity. We randomly selected 40 to 80 percent of the data and discarded the rest to see how our proposed method responds to this problem compared to other methods. With the data obtained from this method, we compared the results of baselines and the proposed method. As shown from the graphs, the proposed method gives a better answer to the data sparsity problem.

\subsection{Effect of social information}
In this experiment, we aim to examine the effectiveness of the social information that we utilize in this model. As such, we compare the performance of our \modelname with its variant, called \modelname-NoSocial. The difference between the two models is on the usage of social information. \modelname-NoSocial does not take social information into account and is based solely on geographical and temporal information. As we see in Table~\ref{tbl:results}, \modelname consistently outperforms \modelname-NoSocial for all evaluation metrics, on both datasets. This indicates the effectiveness of our proposed social model. Moreover, we see in Figures \ref{fig:sparsity-gowalla} and \ref{fig:sparcity-yelp} that \modelname consistently outperforms \modelname-NoSocial as the number of training check-ins decreases, suggesting that the additional social information improves the model's robustness for highly sparse datasets.

\subsection{Effect of model parameter}
\label{sec:param_effect}
\begin{figure}[!tbp]
  \centering
  \subfloat[$\beta$ on Rec@20]{\includegraphics[width=0.46\textwidth]{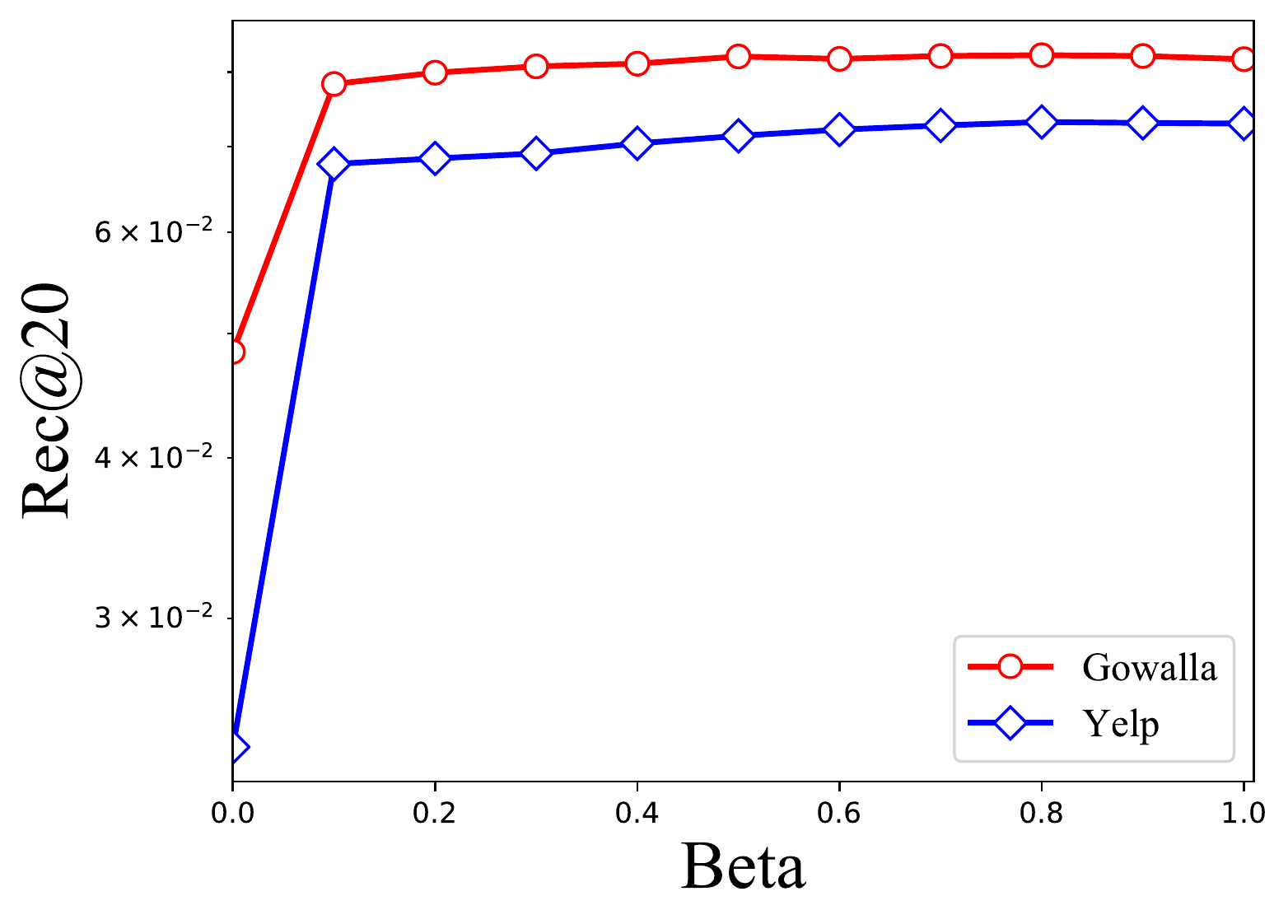}\label{fig:beta-rec}}
  \hfill
  \subfloat[$\beta$ on nDCG@20]{\includegraphics[width=0.46\textwidth]{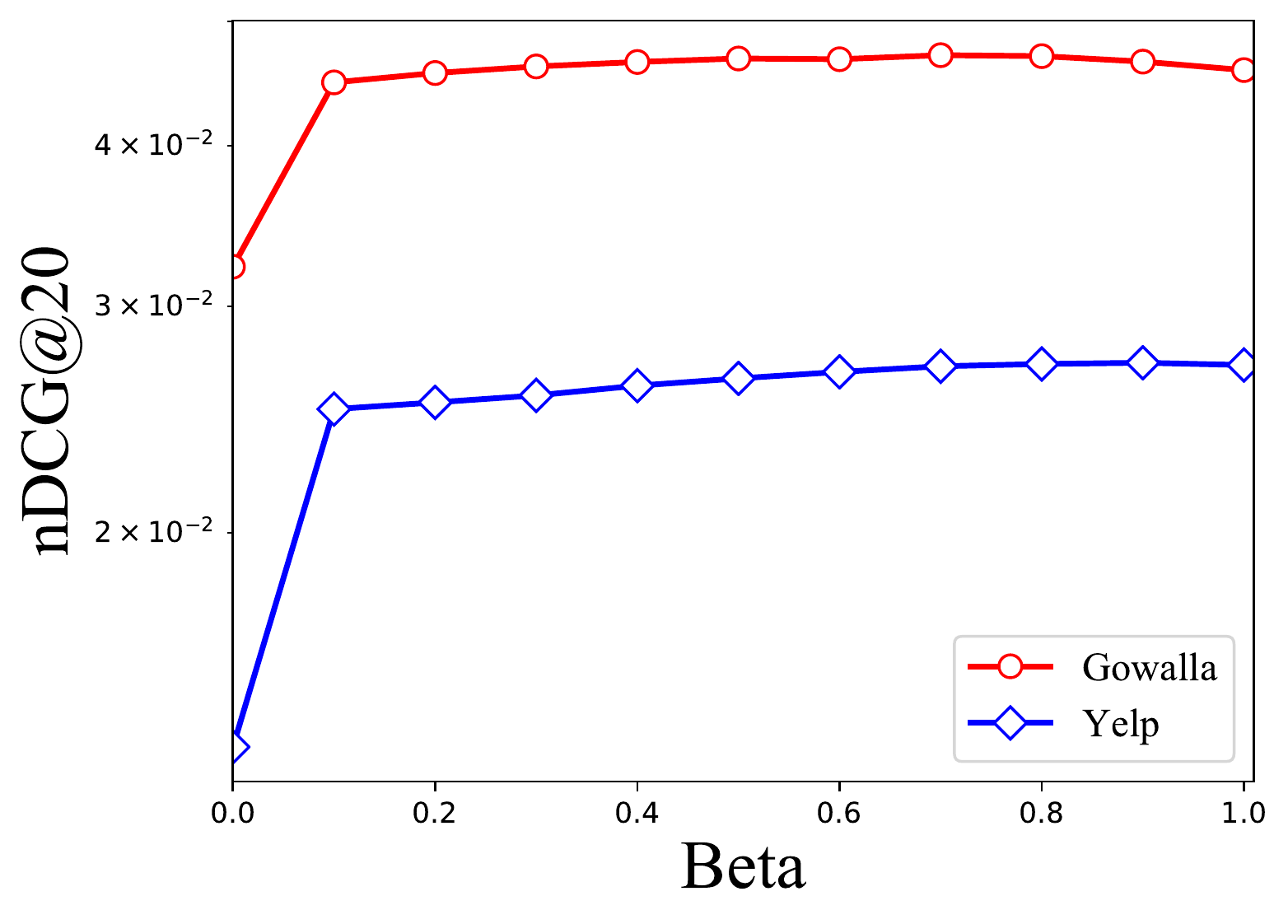}\label{fig:beta-ndcg}}
  \caption{Effect of the $\beta$ on the (a) Rec@20 and (b) nDCG@20 of the performance of \modelname on Gowalla and Yelp.}
  \label{fig:parameters}
\end{figure}

Figure \ref{fig:parameters} shows the performance of \modelname for different values of $\beta$. 
To do so, we train our model using the tune set for different values of $\beta$ in the range of $0.0$ and $1.0$ while keeping all other parameters fixed.
We report in Figure \ref{fig:beta-rec} and \ref{fig:beta-ndcg} the effect of different values of $\beta$ on the performance of \modelname in terms of Recall@20 and nDCG@20, respectively. The plots show that on Gowalla dataset with $\beta=0.7$ and on Yelp dataset with $\beta=0.8$ \modelname achieves the best performance. Therefore, these results indicate that both similarity and friendship edges are important to construct the user and POIs graph. It means that depending on the dataset, similar users based on the commonly visited location, have more impact on the user's visit of new POIs and the performance of the recommendation system.

\section{Discussion} 
\label{sec:discussion}
Our proposed model outperforms LRT and LMFT. These models incorporate temporal information to improve the accuracy of POI recommendation systems. This suggests that incorporating various contextual information helps the model address the sparsity problem more effectively. Moreover, observing Table \ref{tbl:results} more closely we see that SUCP beats the LFBCA model. Given that LFBCA leverages both geographical and social information in the recommendation, it suggests that our proposed method is more effective in incorporating and combining these two sources of information. In particular, related research \cite{wang2013location} has revealed that user check-ins are mostly around different centers and consider that recommendation improves performance. Our proposed model does leverage the user's activity centers and uses this information to locate their friends. This comparison clearly reveals the effectiveness and superiority of this approach.

As mentioned earlier, one of the challenges of recommender systems is data sparsity, which affects system performance. Using more contextual information, the system can have a better view of users' tastes and address the data sparsity problem.
We evaluate the effect of data sparsity on our proposed model. According to the results obtained in Figures \ref{fig:sparsity-gowalla} and \ref{fig:sparcity-yelp}, our model exhibits a more robust performance of various levels of sparsity.
Although the GeoSoCa model uses three different contextual information (i.e., categorical, social, and geographic), In their social model, they use a value of $1$ in the Matrix Factorization to indicate friendship between people, otherwise, the value is given zero. In fact, they only consider POIs that are not visited by the target user based on the aggregation of check-in frequency. Despite the fact that our model integrates friendships and similarities of users. \modelname shows that users who do not have direct friendships but have made joint visits have the potential for friendship and social impact between them.
The L-WMF model provides negative samples based on check-in frequency of users' visit model geographical information by using a neighborhood-aware weight Matrix Factorization. Our model, on the other hand, captures users' activity centers, which can show POIs that are interesting to visit for each user. Furthermore, \modelname incorporates the social information lending to more diversity of the recommended POIs.

The \modelname model outperforms CARA on both datasets significantly. CARA models the users and locations relation based on recurrent neural network architecture and these results indicate the size of the data sets or information is one of the most important factors in the performance of the neural network-based approach. Also, in CARA, the focus is on temporal information more than the other contexts while in \modelname we model social influence based on the time of user's activity in different geographical centers.

\section{Conclusions and Future Work}
\label{sec:conclusion}
POI recommender systems are important tools that are welcomed in LBSNs and business organizations because they increase their popularity and profitability. Although many methods have been developed for POI recommender systems, there are still shortcomings in these systems that require continued study in this area to resolve them. In this paper, we examined the characteristics of user mobility behavior in two large-scale datasets, Yelp and Gowalla, by studying various features related to social information, user's activity center, and check-in time. We analyzed the distance between user-friends activity centers, giving a clearer picture of their behavior and habits. Finally, we proposed the \modelname model as a uniform framework that considers social influence based on user's activity centers and combines them to recommend POIs. Our experimental results on both datasets indicate that \modelname model significantly outperforms the compared and state-of-the-art methods. As for future work, one can be incorporating the other contextual information into the proposed model such as user's comments or reviews, categorical information. Moreover, we may model the social influence by the other friendship and similarity metrics like time of commonly visited check-ins.

\bibliography{mybibfile}

\end{document}